\documentstyle[12pt]{article}

\begin{document} 

\begin{titlepage}

\hrule 
\leftline{}
\leftline{Preprint
          \hfill   \hbox{\bf CHIBA-EP-96}}
\leftline{\hfill   \hbox{\bf OUTP-96-50P}}
\leftline{\hfill   \hbox{hep-ph/9608402}}
\leftline{\hfill   \hbox{August 1996}}
\vskip 5pt
\hrule 
\vskip 1.5cm

\centerline{\large\bf 
Running Coupling Constant of a Gauge Theory 
} 
\centerline{\large\bf  
in the Framework of the Schwinger-Dyson Equation:
}
\centerline{\large\bf
Infrared Behavior of Three-Dimensional 
} \centerline{\large\bf 
Quantum Electrodynamics
$^*$}
   
\vskip 1cm

\centerline{{\bf 
Kei-Ichi Kondo$^{\dagger}$
}}  
\vskip 4mm
\begin{description}
\item[]{\it  
  Theoretical Physics, 
  University of Oxford,
  1 Keble Road, Oxford, OX1 3NP, UK.$^\ddagger$
  }
\item[]{$^\dagger$ 
  E-mail:   kondo@thphys.ox.ac.uk; 
  kondo@cuphd.nd.chiba-u.ac.jp 
}    
\end{description}
\vskip 1cm

\centerline{{\bf Abstract}} \vskip .5cm
We discuss how to define and obtain the running coupling
of a gauge theory in the approach of the Schwinger-Dyson
equation, in order to perform a non-perturbative study
of the theory.  For this purpose, we introduce the
nonlocally generalized gauge fixing into the SD equation,
which is used to define the running coupling constant
(this method is applicable only to a gauge theory).  
Some advantages and validity of this approach are
exemplified in QED3.
This confirms the slowing down of the
rate of decrease of the running coupling and the
existence of non-trivial infra-red fixed point (in
the normal phase) of QED3, claimed recently by Aitchison
and Mavromatos, without so many of their approximations. 
We also argue that the conventional approach is recovered
by applying the (inverse) Landau-Khalatnikov
transformation to the nonlocal gauge result.

\vskip 0.5cm
Key words: Schwinger-Dyson equation, nonlocal gauge,
quantum electrodynamics, running coupling,
renormalization group, fixed point

PACS: 11.15-q, 12.20.-m, 11.10Kk, 11.15.Tk
\vskip 0.5cm
\hrule  


\begin{description}
\item[]{
$^\ddagger$
Address from March 1996 to December 1996.
  On leave of absence from: \\
  Department of Physics, Faculty of Science,
  Chiba University, Chiba 263, Japan.
  }
\end{description}

\end{titlepage}

\pagenumbering{arabic}

\newpage
\section{Introduction}
\setcounter{equation}{0}

In quantum field theory, a change in the renormalization
scale $\mu$ accompanied by suitable change in coupling
and mass does not change the theory. The invariance of
the theory under such transformation is called the
renormalization group (RG) invariance. For a given
physical theory, we have, for each value of
$\mu$, a definite value of the coupling $g(\mu)$ and
$m(\mu)$.  These are called the {\it effective} (or
{\it running}) coupling and mass.
Physical quantities like the S-matrix are invariant under
the change of variable:
$
(\mu, g(\mu), m(\mu)) \rightarrow 
(\mu', g(\mu'), m(\mu')) .
$
This invariance leads to a differential equation for
the running coupling and mass, called the RG equation
\cite{SP53}.
This is the easiest way in practice to
compute the effective coupling and mass.
The coefficients $\beta, \gamma_m, ...$ in the RG equation
are called the RG coefficients which express important
properties of a theory.

\par
The most important application of the RG is to compute
large-momentum, i.e. ultraviolet (UV) behavior.  However,
RG methods can be used to compute infrared (IR)
behavior too.  Certainly this is true in a purely
massless theory (or if masses can be neglected), and the
IR behavior is computable perturbatively if and only if
the theory is not asymptotic free. But in a massive
theory it is not useful to take $\mu$ much less than a
typical mass, for one obtains logarithms of $m/\mu$ and
these prevent simple use of the perturbative method when
$\mu
\ll m$, see e.g.
\cite{Collins84}.
\par
In four spacetime dimensions, it is well known that the
only theories that are asymptotically free are non-Abelian
gauge theories with a small enough number of matter
fields, see
\cite{CG73}.
In an asymptotically free theory, the effective coupling
$g(\mu)$ goes to zero when $\mu$ goes to infinity, so
that short-distance (or high-energy) behavior is
computable perturbatively.  But, when $\mu$ is small,
$g(\mu)$ is large, so that IR behavior cannot be
computed reliably by perturbation theory.  This is the
case for QCD$_4$.
\par
If we consider a non-asymptotically free theory in four
dimensions, like
$(\phi^4)_4$ and QED$_4$, then the effective coupling
increases with energy.  Thus, in such theories it is
impossible to compute the true high-energy behavior by
perturbation theory.
But  when $\mu$ goes to zero, so does the effective
coupling.  Hence, we can compute IR behavior in such a
theory, just as we compute the UV behavior in an
asymptotically free theory.  
However, note that the coupling in QED is 
$\alpha :=e^2/4\pi \sim 1/137$.  This is so small that
the non-perturbative region in QED$_4$ does not occur
until very many orders of magnitude beyond
experimentally accessible energies.
Nevertheless, the existence of a non-trivial UV fixed
point in a non-asymptotically free theory in four
dimensions has been suggested by
Miransky \cite{Miransky85} for strongly coupled QED$_4$
where the fixed point is expected to be located at
$\alpha=\alpha_* \sim O(1)$ of order unity. In the strong
coupling region, we need to deal with the theory
non-perturbatively.
\par
As well as simulations based on lattice gauge theory
\cite{Kogut88},  the Schwinger-Dyson (SD) equation has
played an important role in the non-perturbative
analytical study of strongly coupled gauge theories.
In the actual analysis of the SD equation, an
approximation of constant coupling (standing coupling)
has been taken together with a bare vertex
approximation as the simplest approximation, which is
usually called the quenched planar (or ladder)
approximation. However, the introduction of running
coupling is indispensable to study the unquenched QED,
QCD and extended technicolor theory, etc, see e.g.
\cite{SCGT91}. So far, this has been done in most cases
in a somewhat unsatisfactory way, in my opinion, in which
the expression for the running coupling constant obtained
by (RG-improved) perturbation theory (at most leading
logarithm) was substituted into the SD equation.
However, the running coupling itself should be calculated
within the SD equation approach.  Such a kind
of calculation was tried, for example, in the massive
gauge boson theory in four-dimensions \cite{Kondo89} and
$N$-flavor QED$_3$ \cite{PW88,Nash89,AJM90,KN92,BR93}
by solving the SD equation for the wavefunction
renormalization function
$A(p^2)$ of the fermion, using either a bare vertex or a
reasonably simple ansatz for the vertex function. 
In the original analysis \cite{Pisarski84,ABKW86} of
multi-flavor QED$_3$, no wavefunction renormalization was
assumed from the beginning, based on a naive $1/N$
argument.  Quite recently, self-consistent solutions of
QED$_3$ have been studied more extensively by solving the
coupled SD equation under various ansatzes for the vertex
\cite{Maris96}.
 
\par
The choice of vertex ansatz is the most difficult problem
in truncating the infinite hierarchy in the SD
equation approach.
Even if the vertex ansatz might be reasonable, 
we need to make a number of approximations in solving the
SD (integral) equation, at least analytically.  The nature
of those approximations is totally different from what one
encounters in solving a differential equation. Usually,
obtaining an analytical solution for an integral
equation is much more difficult than for a differential
equation. Therefore, the SD integral
equation is often converted into a differential equation
by simplifying the kernel, although the integral equation
can not in general be transformed into a differential
equation in a mathematically rigorous sense. 
\footnote{
The numerical approach enables us to
solve the integral equation as well as the differential
equation, although the accuracy and the convergence
of the algorithm for solving the integral equation has not
been established so well. 
Justification of the approximations made in the
analytical study has been done through the numerical
calculations.
}

\par
In this paper 
we discuss another approach to calculating the running or
effective coupling in a gauge theory.  In this approach
we introduce the nonlocal gauge-fixing
\cite{LK56,GSC90,KM92}. Here the existence of gauge
degrees of freedom is an essential ingredient, so this
approach is applicable to a gauge theory only.  
\par
\par
The conventional approach relies on a specific
choice of vertex function.
Therefore, one can not be free from the criticism
whether the result is due to an artifact of the
specific ansatz adopted for the vertex function or not.
Of course, it is impossible to get rid of this criticism
completely. However, we can allow a certain class of
ansatz for the vertex function which gives a weaker
restriction than the conventional approach.
In this paper, we choose the ansatz for the vertex:
\begin{equation}
\Gamma_\mu(p,q) = \gamma_\mu G(p^2, q^2, k^2)
\end{equation}
where the function $G$ is an {\it arbitrary} function of
the fermion momenta $p^2$, $q^2=(p-k)^2$ and the
gauge-boson momentum $k^2$.
 We shall see that the resulting running coupling is
essentially independent
of the explicit form of $G$, provided the vertex
function satisfies the Ward-Tahakashi identity.
This is a major advantage of this approach.
\par
So far, the nonlocal gauge has been applied to gauge
theories in four dimensions \cite{GSC90,KM92}, the
gauged Thirring model \cite{Kondo96a}
and QED$_3$ \cite{Simmons90,KEIT95,KM95}.  In those
works, a bare vertex approximation has been taken from the
beginning.  This assumption is not necessary or is
weakened as shown in this paper.

\par
The second advantage of this approach is as follows.
This approach is free from various approximations which
are required to solve the SD equation analytically for the
wavefunction renormalization, because
we never solve it to obtain the running
coupling constant. Instead, we derive a differential
equation which should be satisfied by the nonlocal gauge
function. 
The nonlocal gauge is obtained by a simple quadrature
without solving any self-consistent equation.
This is an
advantage of this approach.
\par
We apply this method to study the non-perturbative
behavior of QED$_3$, in particular, the IR behavior and
the rate of running in the intermediate momentum region
which have been extensively studied recently 
by Aitchison and Mavromatos \cite{AM96} and their
collaborators \cite{AAKMN96}. We also present the RG-like
argument for the behavior of the running coupling within
this approach.
\par
This paper is organized as follows.
In section 2, we introduce the nonlocal gauge fixing and
set up the SD equation under a class of ansatzes for the
vertex function in
$D(\ge 2)$-dimensional gauge theory.
\par
In section 3, we derive the differential equation which
is obeyed by the nonlocal gauge such that there is no
wavefunction renormalization $A(p^2) \equiv 1$.
Here we introduce the nonlocal gauge with and without IR
(not UV) cutoff.  The IR cutoff is necessary to discuss
the IR behavior of QED$_3$ in the following sections.
\par
In section 4, we discuss how to obtain the non-trivial
wavefunction renormalization in the usual (Landau) gauge
from the result in the nonlocal gauge.  For this, we
apply the inverse LK transformation.
\par
In section 5, we define the running coupling through the
nonlocal gauge in this approach and compare it with
that in the conventional approach.  Here we see
several advantages of this approach for obtaining the
running coupling.
\par
In section 6, we study in detail the behavior of the
running coupling of QED$_3$ obtained in the previous
section, paying particular attention to the IR behavior.
The RG-like interpretation of this result is also
given here.
\par
The final section is devoted to conclusion and
discussion.

\section{Schwinger-Dyson equation}
\setcounter{equation}{0}

\subsection{Introduction of the nonlocal gauge}

First, we discuss the SD equation for the fermion
propagator.   
In accord with the bare fermion propagator 
\begin{equation}
  S_0(p) = (\gamma^\mu p_\mu - m_0)^{-1},
\end{equation}
we write the full fermion propagator as
\begin{equation}
  S(p) = [A(p^2) \gamma^\mu p_\mu  - B(p^2)]^{-1} .
\end{equation}
For a class of gauge theories in $D=d+1$ dimensional
space-time, the SD equation for the full fermion
propagator in momentum space is given by
\begin{equation}
  S^{-1}(p) = S_0^{-1}(p) + \int {d^Dq \over (2\pi)^D}
  \gamma_\mu S(q) \Gamma_\nu(p,q) D_{\mu\nu}(p-q),
  \label{SD}
\end{equation}
where $\Gamma_\nu(p,q)$ is the full vertex function 
and $D_{\mu\nu}(p-q)$ is the full gauge boson
propagator. We always use $p, q$ for the fermion
momentum and $k=p-q$ for the gauge-boson momentum.  
This class includes QED and the gauged Thirring model
\cite{Kondo96a,Kondo96c}. 
Note that this SD equation can be decomposed
into a pair of integral equations for the wavefunction
renormalization function
$A(p^2)$ and the mass function
$B(p^2)$, as shown in the next section.
The SD equation for the full fermion propagator $S(p)$
should constitute a closed set of equations together
with the SD equations for the full vertex function
$\Gamma_\nu(p,q)$ and the full gauge boson propagator
$D_{\mu\nu}(p-q)$ which will be specified below.

\par
Next, in order to specify the gauge boson propagator, we
discuss the gauge-fixing.
In this paper we consider a more general
gauge fixing than the usual one, the so-called the
{\it nonlocal} gauge-fixing. In configuration space,
the gauge fixing term in the nonlocal gauge \cite{KM95} is
given by
\begin{equation}
  {\cal L}_{GF}
  = - {1 \over 2} F[A(x)] \int d^D y
  {1 \over \xi(x-y)}F[A(y)],
\label{nlgfterm}
\end{equation}
with a gauge-fixing function $F[A]$.  In this paper we
take the Lorentz-covariant linear gauge:
\begin{equation}
F[A]=\partial^\mu A_\mu .
\label{covgauge}
\end{equation}
In momentum representation, the gauge-fixing parameter
$\xi$ becomes momentum-dependent, namely,
$\xi$ becomes a function of the momentum:
$\xi=\xi(k)$.
Here it should be noted that  $\xi^{-1}(k)$ is the
Fourier transform of $\xi^{-1}(x)$: 
\begin{equation}
  \xi^{-1}(x) = \int {d^Dk \over (2\pi)^D}
e^{ikx} \xi^{-1}(k),
\quad
  \xi^{-1}(k) = \int  d^D x
e^{-ikx} \xi^{-1}(x),
\end{equation}
while $\xi(k)$ is not the Fourier
transform of $\xi(x)$.
If $\xi(k)$ does not have  
momentum-dependence, i.e.,
$\xi(k) \rightarrow \xi$, then 
$\xi^{-1}(x-y) \rightarrow \delta(x-y)\xi^{-1}$ and
hence the nonlocal gauge-fixing term (\ref{nlgfterm})
reduces to the usual gauge-fixing term: 
\begin{equation}
  {\cal L}_{GF}
  = - {1 \over 2\xi} (F[A(x)])^2 .
\label{gfterm}
\end{equation}
\par
It is easy to show that the SD equation for the full
gauge-boson propagator is given by
\begin{eqnarray}
 D_{\mu\nu}^{-1}(k) 
 &=& D^{(0)}_{\mu\nu}{}^{-1}(k) - \Pi_{\mu\nu}(k),
 \nonumber\\
 \Pi_{\mu\nu}(k) &:=& e^2 \int {d^Dp \over (2\pi)^D}
 {\rm tr}[\gamma_\mu S(p) \Gamma_\nu(p,p-k) S(p-k)],
\end{eqnarray}
where the bare gauge-boson propagator 
$D^{(0)}_{\mu\nu}(k)$ in the nonlocal
gauge (\ref{covgauge}) is given by
\begin{eqnarray}
 D^{(0)}_{\mu\nu}{}^{-1}(k) 
 = k^2 g_{\mu\nu} -  k_\mu k_\nu +  \xi(k)^{-1} k_\mu
k_\nu .
\end{eqnarray}
In gauge theory, the vacuum polarization tensor
should have  the transverse form:
\begin{eqnarray}
 \Pi_{\mu\nu}(k)  
 =  \left( g_{\mu\nu} - {k_\mu k_\nu \over k^2} \right) 
 \Pi(k),
\end{eqnarray}
provided the gauge invariance is preserved.
Hence the full gauge-boson propagator is of the form
\begin{eqnarray}
 D_{\mu\nu}(k) 
 &=& D_T(k) \left( g_{\mu\nu} - {k_\mu k_\nu \over k^2}
\right)
 + {\xi(k) \over k^2} {k_\mu k_\nu \over k^2},
 \nonumber\\
 D_T(k) &:=& {1 \over k^2 - \Pi(k)} .
 \label{gbpropa}
\end{eqnarray}
\par
In this paper we take another form for
the full gauge boson propagator:
\begin{equation}
  D_{\mu\nu}(k) 
  = D_T(k^2)
    \left[ g_{\mu\nu} - \eta(k^2) \frac{k_\mu k_\nu}{k^2}
\right] , 
\quad k_\mu := p_\mu - q_\mu.
\label{gbpropanl}
\end{equation}
By comparing (\ref{gbpropa}) with (\ref{gbpropanl}),
the correspondence between $\xi$ and
$\eta$ is given in momentum space as follows.
\begin{eqnarray}
  \xi(k) = [1-\eta(k^2)][1 -\Pi(k)/k^2]^{-1},
  \quad
  \eta(k^2) = 1-\xi(k)[1 - \Pi(k)/k^2] .
\label{gaugerel}
\end{eqnarray} 

\subsection{Vertex ansatz and gauge choice}
\par
Finally, we must specify the full vertex function
$\Gamma_\nu(p,q)$.  Of course, the vertex function obeys
its own SD equation and all the SD equations should be
solved simultaneously.  In principle, it might be possible
to do that.  However, it is rather difficult or
impossible to actually carry out this scenario, except
for some exactly solvable models (see
e.g.\cite{Kondo96c}), due to the infinite hierarchy of the
SD equations.   Therefore we need to truncate the SD
equations so that they become tractable
analytically or numerically.
One usually adopt an ansatz for
the full vertex function $\Gamma_\mu$, instead of solving the SD equation
for the vertex function.  Such an ansatz is
suggested from various consistency requirements, 
see e.g. \cite{Pennington95}.
In this paper we adopt the following simple ansatz:
\begin{eqnarray}
 \Gamma_\mu(p,q) = G(p^2, q^2, k^2) \gamma_\mu,
 \label{vertex1}
\end{eqnarray}
where $p_\mu, q_\mu$ denotes the momenta of the
fermion and $k_\mu$ the momentum of the gauge boson. 
Here $G=G(p^2, q^2, k^2)$ is an arbitrary function of
$p^2, q^2$ and $k^2$ except for a restriction specified
below. 
\par
The truncation of the set of SD equations in
the gauge theory should be done self-consistently in such a
way that the resulting truncated SD equations  respect 
gauge invariance (as much as possible).   For example, the
vertex function should satisfy the Ward-Takahashi (WT)
identity:
\begin{eqnarray}
  (p-q)^\mu \Gamma_\mu(p,q)  =  S(p)^{-1} - S(q)^{-1},
  \label{WT}
\end{eqnarray}
which is a consequence of the gauge invariance of the
theory. When the function $B(p^2)$ vanishes (in the
symmetric phase or normal phase) or is extremely small (in
the neighborhood of the critical point in the broken phase
or superconducting phase), the WT identity reduces to
\begin{eqnarray}
  (p-q)^\mu \Gamma_\mu(p,q)  
  \sim   [A(p^2) p^\mu - A(q^2) q^\mu] \gamma_\mu .
  \label{WT2}
\end{eqnarray}
This tells us that $G(p^2, q^2, k^2)$ should be expressed
in terms of $A(p^2), A(q^2)$ and $A(k^2)$:
\begin{eqnarray}
  G(p^2, q^2, k^2) 
  = F[p^2, q^2, k^2, A(p^2), A(q^2), A(q^2)],
\end{eqnarray}
Some examples will be shown below.
A restriction for our approach to work is
\begin{eqnarray}
  {\partial G(p^2, q^2, k^2) \over \partial k^2}
  \Biggr|_{p^2,q^2}
  = 0,
  \label{restriction}
\end{eqnarray}
where the derivative is taken with $p^2$ and $q^2$ being
fixed.
This implies that the argument of $G$ does not depend
{\it explicitly} on the angle $\vartheta$ of the inner
product 
$p \cdot q=pq \cos \vartheta$ which comes from 
$k^2 := (p-q)^2 = p^2+q^2-2p \cdot q$.
However, $G$ can depend on $k^2$ implicitly, through
$A(k^2)$.
The reason why we need this restriction will be
made clear in the next section.
In what follows we restrict our discussion to this case.
\par
The simplest choice is the bare vertex approximation:
$
 \Gamma_\mu(p,q)  \equiv  \gamma_\mu. 
$
This approximation is usually called the ladder
approximation.
\footnote{
This naming is somewhat misleading, since  the SD
equation for the fermion propagator does not have
an exact graphical representation corresponding to  
ladders.  
}
And it is said that the ladder approximation
breaks gauge invariance 
(this is obvious, since (\ref{WT2}) can not be
satisfied unless
$A(p^2) \equiv 1$ which is impossible as a solution of
the SD equation except Landau gauge.).   This
statement is sometimes used as indicating that the ladder
approximation in the SD equation may lead to wrong result
in the analysis of the gauge theory.  Here we define the
{\it ladder} approximation in the SD equation (for the
fermion propagator) as an approximation in which the full
vertex function is replaced with the bare one. In our
definition of ladder approximation, the gauge boson
propagator is not necessarily restricted to the bare one. 
\par
However, in our approach we look for a set of consistent
solutions for 
$A(p^2)$ and $B(p^2)$ together with the
vertex function $\Gamma_\mu(p,q)$ and the gauge boson
propagator $D_{\mu\nu}(k)$ under the adopted
approximation, so that the solutions are consistent with the
WT identities.  
For the bare vertex approximation to be consistent with
the WT identity (\ref{WT2}), therefore, there should be no
wavefunction renormalization for the fermion. In other
words, the bare vertex approximation should
self-consistently yield the result: 
\begin{equation}
 A(p^2) \equiv 1, 
 \label{nowfr}
\end{equation}
as a solution of SD equation Eq.~(\ref{SD}).  
In QED with covariant gauge (\ref{gfterm}),
(\ref{nowfr}) is realized only if we take both the
Landau gauge and the quenched ladder approximation where
the full photon propagator is also replaced with the bare
one, i.e. $\Pi(k) \equiv 0$.   
\footnote{
In the usual perturbation theory of QED,
the Landau gauge is a special gauge in which the vertex
correction vanishes.   The above example shows that the
similar situation occurs also in the framework of the SD
equation.   
}
Therefore, within the framework of SD equations in QED,
the bare photon propagator in the Landau gauge together
with the bare vertex is a set of consistent solutions of
the SD equation (the fermion propagator can be
non-trivial).
\footnote{
It is argued \cite{KITE94} that the quenched
limit of gauge-invariant study of QED$_4$
\cite{Fukuda} coincides with the Landau-gauge results
obtained from the Schwinger-Dyson equation in the quenched
ladder (bare vertex) approximation.
This suggest that the Landau gauge is the best one in this
approximation. }  
This fact is well-known in four dimensions \cite{CG89} and
holds in any dimension $D>2$, see \cite{KN89}.   
\par
If gauges other than the Landau gauge are adopted,
$A(p^2) \equiv 1$ does not hold even in
the quenched ladder (bare vertex and bare gauge boson
propagator) approximation and we have to solve the
coupled equations for
$A$ and $B$.  However, it is known that the solutions in
QED obtained in such a scheme are severely gauge-dependent,
see e.g. \cite{Haymaker82}. 
Indeed, the simplest quenched ladder
approximation does not satisfy the WT identity except for the
Landau gauge in the sense described above.  In order to
obtain gauge-parameter independent results in this case,
we need to modify the full vertex so as to satisfy the WT
identity.   In the framework of the SD equation, however,
it is rather difficult to obtain gauge-parameter
independent results by modifying the vertex function.
\footnote{
In order for the full vertex function to be consistent
with the Ward-Takahashi identity, the ansatz for the
vertex function should include $S$, e.g.
$\Gamma_\mu(p,q) = {k_\mu \over k^2}[S^{-1}(p)-S^{-1}(q)]
+ \Gamma_\mu^T(p,q)$.
However, this requirement is not sufficient to determine
the vertex uniquely, see ref.
\cite{Pennington91,Kondo92,Atkinson94} and references
therein. 
For an other proposal, see \cite{Kondo96b}.
}
\par
As explained above, in the quenched ladder approximation,
only the Landau gauge
$\eta(k^2) \equiv 1$ can give a set of consistent solutions
for the SD equation in this sense. However, if the vacuum
polarization is included in the photon propagator (i.e.
unquenched case $\Pi(k) \not= 0$), the Landau gauge
has no longer this property in the SD framework. 
Adopting the {\it nonlocal gauge}  enables us to extend
this scheme beyond the quenched ladder approximation.  In
a gauge theory, there is the freedom of choosing such a
gauge.   The existence of a gauge where $A(p^2) \equiv 1$
in the SD framework was shown independently 
by Georgi, Simmons and Cohen \cite{GSC90} and 
by Kugo and Mitchard \cite{KM92} 
in four-dimensional gauge theory. The extension to
arbitrary dimension $D(>2)$ is straightforward, as done
in the Appendix of \cite{KEIT95}. 
There the bare vertex was assumed from the beginning.
In the next section we re-derive the nonlocal gauge in
a more general setting.
\par
As long as the full vertex has the form
(\ref{vertex1}), the nonlocal gauge plays the same role
as the Landau gauge of the quenched ladder QED.   In the
next section, we show that the function
$\eta(k^2)$ can be chosen so that the SD equation
Eq.~(\ref{SD}) for the fermion propagator with the vertex
function (\ref{vertex1}) leads to the solution
Eq.(\ref{nowfr}) for $A(p^2)$, i.e. no wavefunction
renormalization. Therefore, the nonlocal gauge gives the
most economical choice, since we have only to solve the
single equation for $B$.   

\par
So far, we have not discussed the explicit dependence of
$G$ on $A$.
Examples used so far for the ansatz belonging to the
class (\ref{vertex1}) are as follows.
\begin{eqnarray}
G(p^2, q^2, k^2) &=& A(p^2), \quad A(q^2) ,
\nonumber\\
  &=& {1 \over 2}[A(p^2)+A(q^2)], 
\nonumber\\
  &=& A(p^2) \theta(p^2-q^2)
+ A(q^2) \theta(q^2-p^2) ,
\nonumber\\
  &=& 
\left[ A(p^2) + c_1 p^2 A'(p^2) + c_2 p^4 A''(p^2) 
+ ... \right] \theta(p^2-q^2) + 
(p \leftrightarrow q) , ...
\end{eqnarray} 
with some constants $c_i ( i=1,2,...)$.
If we take the nonlocal gauge so that $A(p^2)
\equiv 1$, the function $G(p^2, q^2, k^2)$ reduces to 1
uniformly in
$p, q, k$ (when $B(p^2)$ is neglected), in order to be
consistent with the WT identity. 
The following ansatzes are incompatible with the WT
identity:
\begin{eqnarray}
G(p^2, q^2, k^2) &=& A^n(p^2), \quad A^n(q^2) ,
\nonumber\\
  &=& {1 \over 2^n}[A(p^2)+A(q^2)]^n, 
\nonumber\\
  &=& A(p^2) A(q^2),
\nonumber\\
  &=& A(p^2)A(q^2)/A(k^2), 
\end{eqnarray} 
for any integer $n \ge 2$.

\par
In the massive fermion phase $B(p) \not=0$, the small
deviation of
$G(p^2, q^2, k^2)$ from 1 is
${\cal O}(B)$ which gives rise to at least
${\cal O}(B^2)$ terms in the integrand of the SD equation
for $B$. In order to study the critical behavior in the
neighborhood of the critical point through the solution of
the SD integral equation for
$B(p^2)$, we can put $G(p^2, q^2, k^2) \equiv 1$ in the
integrand of the integral equation after taking the
nonlocal gauge, even if
$G(p^2, q^2, k^2)$ might deviate from 1 off the critical
point.   This is because the order ${\cal O}(B^2)$ terms
are irrelevant to the bifurcation solution (from the
trivial one $B(p^2) \equiv 0$) which is sufficient to study
the critical behavior. 
\par
Therefore, in this
approach we do not need the explicit form of the vertex
in order to study the critical behavior of the model.
This is a further advantage of this approach.  For
other types of the full vertex with different tensor
structure \cite{Pennington91,Kondo92,Atkinson94},
such a convenient gauge is not known and we must solve
the coupled equation for $A$ and
$B$ as well as $\Gamma_\mu$.  
In our approach all effects coming from the vertex
correction are incorporated into the SD equation by
modifying the gauge boson propagator through the nonlocal
gauge.
\par
In QED$_3$ it has been shown \cite{KM95} that the nonlocal
gauge reproduces systematically the previous
result obtained for the fermion self-energy with
corrections up to the next-to-leading order in $1/N$
expansion \cite{Nash89}, but in the approach of
\cite{Nash89} it was necessary to consider the vertex
correction in writing down the SD equation.

\par

\section{Derivation of nonlocal gauge}
\setcounter{equation}{0}
\par
We decompose the SD equation into a pair of
integral equations according to the following procedure:
\begin{eqnarray}
 A(p^2) &=& 1 
 + {{\rm tr}[\Sigma(p^2) \gamma^\mu p_\mu] \over p^2
{\rm tr}(1)},
 \label{decompSDA}
\\
 B(p^2) &=& m_0 
 + {{\rm tr}[\Sigma(p^2)] \over p^2 {\rm tr}(1)},
 \label{decompSDB}
\end{eqnarray} 
where $\Sigma$ denotes the self-energy part:
\begin{equation}
  \Sigma(p^2) := \int {d^Dq \over (2\pi)^D}
  \gamma_\mu S(q) 
  \Gamma_\nu(q,p) D_{\mu\nu}(p-q).
  \label{selfenergy}
\end{equation}
We use the ansatz (\ref{vertex1}) for the
vertex function.
\par
Then the SD equation (\ref{decompSDA}) for the fermion
wave function renormalization $A$ reads
\begin{eqnarray}
 && p^2 A(p^2) -  p^2
\nonumber\\ &=& e^2 \int {d^Dq \over (2\pi)^D} 
{A(q^2) G(p^2, q^2, k^2) \over q^2
A^2(q^2)+B^2(q^2)}
\nonumber\\ &&  \times k^2 D_T(k^2)
\left[ (D-2) {p \cdot q \over k^2} + \left( {p
\cdot q \over k^2} - 2 {p^2q^2-(p \cdot q)^2 \over
k^4} \right) \eta(k^2) \right] .
\label{SDA1}
\end{eqnarray} 
On the other hand, the SD equation (\ref{decompSDB}) for
the fermion mass function $B$ reads
\begin{eqnarray}
 B(p^2) &=& m_0 + e^2 \int {d^Dq \over (2\pi)^D} {B(q^2)
G(p^2, q^2, k^2) \over q^2 A^2(q^2)+B^2(q^2)}
D_T(k^2) [D-\eta(k^2)] .
\label{SDB}
\end{eqnarray} 
\par
Separating the angle $\vartheta$ defined by
\begin{equation}
  k^2 := (q-p)^2 = x + y -2\sqrt{xy} \cos\vartheta,
\  x := p^2, \qquad y := q^2,
\end{equation}
we find (for $D>2$)
\begin{eqnarray}
 && x A(x) -  x
\nonumber\\ &=&   C_D e^2 \int_0^{\Lambda^2} dy 
{y^{(D-2)/2}A(y) \over y A^2(y)+B^2(y)} 
  \int_0^{\pi} d \vartheta \sin ^{D-2} \vartheta   
  G(p^2, q^2, k^2) 
\nonumber\\ && \times k^2 D_T(k^2)  \left\{  {
\cos \vartheta \sqrt{xy}[D-2+\eta(k^2)] \over k^2}
 - 2 {xy-(\sqrt{xy} \cos \vartheta)^2 \over k^4}  
\eta(k^2) \right\},
\end{eqnarray} 
where
\begin{equation}
 C_D := {1 \over 2^D \pi^{(D+1)/2} 
 \Gamma({D-1 \over 2})} .
\end{equation}

\par
 We follow the same procedure as that given in
\cite{KM92} and Appendix of ref.\cite{KEIT95}.
We perform the angular integration by parts according to
\begin{eqnarray}
&&  \int_0^\pi d\vartheta \sin^{D-2} \vartheta \cos\vartheta
\sqrt{xy} f(z)
 \nonumber\\
 &=& \frac{\sqrt{xy}}{D-1} \left[ \sin^{D-1} \vartheta
f(z)
\right]_0^\pi
   -\frac{2xy}{D-1} \int_0^\pi d\vartheta \sin^D
\vartheta {\partial \over \partial z} f (z),
\end{eqnarray}
where 
$
 z := k^2 = (q-p)^2 
$
and the differential with respect to $z$ is done with
$x$ and $y$ being fixed. 
Thus we find
\begin{eqnarray}
 && x A(x) -  x
\nonumber\\ &=&  
 - {C_D e^2 \over D-1} \int_0^{\Lambda^2} dy
{y ^{(D-2)/2}A(y) (2xy) \over y A^2(y)+B^2(y)} 
 \int_0^{\pi} d \vartheta \sin ^{D} \vartheta 
\nonumber\\ &&   \times
\left[  {\partial \over \partial z} 
\left\{ G(x, y, z) [D-2+\eta(z)]D_T(z)  \right\}
 + (D-1) {G(x, y, z) D_T(z) \eta(z) \over z}   
\right]   .
\end{eqnarray} 
 This is further rewritten as
\begin{eqnarray}
 && x A(x) -  x
\nonumber\\ &=&   - {C_D e^2 \over D-1}
\int_0^{\Lambda^2} dy 
{y^{(D-2)/2}A(y) (2xy) \over y A^2(y)+B^2(y)}  
  \int_0^{\pi} d \vartheta \sin ^{D} \vartheta   
\nonumber\\ && \times 
  {1 \over z^{D-1}}
\Biggr[ G(x,y, z) \left\{  
{\partial \over \partial z}[z^{D-1}  D_T(z) \eta(z)]  
+ (D-2) z^{D-1} {\partial \over \partial z}D_T(z) \right\} 
\nonumber\\ && \quad \quad \quad + 
{\partial G(x,y, z) \over \partial z}
  [D-2+\eta(z)] z^{D-1} D_T(z) \Biggr] .
\end{eqnarray} 
 From this equation, it turns out that the
requirement
$A(p^2) \equiv 1$ is achieved irrespective of $B$, if 
\begin{eqnarray}
{\partial G(x,y, z) \over \partial z }= 0
\end{eqnarray}
and
$\eta(k^2)$ satisfies the following differential
equation:
\begin{eqnarray}
 {\partial \over \partial z}[z^{D-1} D_T(z) \eta(z)] =
- (D-2) z^{D-1} {\partial \over \partial z}D_T(z) .
 \label{de}
\end{eqnarray}

\par
Thus, once the function $D_T(k^2)$ is given, we can find the
nonlocal gauge $\eta(k^2)$ by solving Eq.~(\ref{de}), so
that
$A(k^2) \equiv 1$ follows under the ansatz for the vertex
function  (\ref{vertex1}), provided that $G$ does not
depend {\it explicitly} on
$k^2$. 
Then we have only to solve Eq.~(\ref{SDB}) for the
fermion mass function
$B(p^2)$.
\begin{eqnarray}
 B(p^2) &=& m_0 + e^2 \int {d^Dq \over (2\pi)^D} {B(q^2)
\over q^2 +B^2(q^2)} G(p^2, q^2, k^2) D_T(k^2)
[D-\eta(k^2)] ,
\label{SDBnlg}
\end{eqnarray} 
or
\begin{eqnarray}
 B(x) = m_0 + C_D e^2  \int_0^{\Lambda^2} dy 
 {y^{(D-2)/2}B(y) \over y+B^2(y)} G(x, y, z) K(x,y),
 \label{SDeqB}
\end{eqnarray} 
where the kernel is given by
\begin{eqnarray}
K(x,y) := \int_0^{\pi} d \vartheta \sin ^{D-2}
\vartheta   D_T(k^2) [D-\eta(k^2)] .
\label{kernel}
\end{eqnarray} 
\par
\par
The quenched ladder QED in the covariant gauge is recovered
by taking
$\Pi(k) \equiv 0$, i.e. $D_T(z)=1/z$.  In this case
the nonlocal gauge reduces to the local gauge: 
$\eta(k^2) \equiv \eta = 1-\xi$.  This reproduces the
well-known result: 
$A(p^2) \equiv 1$ in the gauge $\eta = 1$, i.e. 
Landau gauge $\xi = 0$ for $D>2$. 
Especially, for $D=2$, $A(x) \equiv 1$ is satisfied in the
gauge
$\eta \equiv 0$, i.e.  Feynman gauge $\xi = 1$, as can
be seen from Eq.~(\ref{SDA1}).
However, in $D=2$, the SD equation can be exactly solved
in an arbitrary gauge, see \cite{Kondo96c}.

\subsection{Nonlocal gauge without IR cutoff}

The differential equation (\ref{de}) is a first
order differential equation and is simply solved by
choosing a boundary condition.
Integrating both sides of Eq.~(\ref{de}) from $0$
to $k^2$, we obtain
\begin{eqnarray}
 \eta(k^2) &=& - {D-2 \over (k^2)^{D-1} D_T(k^2)}
\int_0^{k^2} dz D_T'(z) z^{D-1} ,
\label{nlg1}
\end{eqnarray} 
where the prime denotes the differential with respect to
$z$. 
Here we have assumed the boundary condition:
\begin{eqnarray}
[z^{D-1} D_T(z) \eta(z)]|_{z=0}=0,
\end{eqnarray} 
so as to eliminate the $1/z^{D-1}$ singularity in
$\eta(z)$. 
Alternatively, we can write
\begin{eqnarray}
 \eta(z) =  (D-2) \left[
{D-1 \over z^{D-1} D_T(z)} \int_0^z dt D_T(t) t^{D-2} -1
\right],
\label{nlg2}
\end{eqnarray} 
where we have assumed that 
\begin{eqnarray}
[z^{D-1} D_T(z)]|_{z=0}=0.
\end{eqnarray} 
 This should be checked after having obtained
the function
$\eta(z)$.

\subsection{Nonlocal gauge with IR cutoff}

For later convenience, we introduce an IR cutoff
$\epsilon$ in the nonlocal gauge by integrating 
(\ref{de}) from
$\epsilon^2$ to $k^2$: 
\begin{eqnarray}
 \eta(k^2) &=& 
 {\epsilon^{2(D-1)} D_T(\epsilon^2) \over
 (k^2)^{D-1} D_T(k^2)} \eta(\epsilon^2)
 - {D-2 \over (k^2)^{D-1} D_T(k^2)}
\int_{\epsilon^2}^{k^2} dz D_T'(z) z^{D-1} ,
\label{nlgir0}
\end{eqnarray} 
where $\eta(\epsilon^2)$ is undetermined.
Note that Eq.~(\ref{de}) is rewritten as
\begin{eqnarray}
 \eta(z) = - {z^{D-1}D_T(z) \eta'(z) + (D-2)z^{D-1}D_T'(z)
 \over [z^{D-1} D_T(z)]'} .
\end{eqnarray}
Therefore, if we impose the flatness condition 
$\eta'(\epsilon^2)=0$ on $\eta(k^2)$ at $k^2=\epsilon^2$
as a boundary condition,
\footnote{
This means the flatness of the effective coupling, see
section 6.
}
the value
$\eta(\epsilon^2)$ is determined as
\begin{eqnarray}
 \eta(\epsilon^2) = -
{(D-2) (\epsilon^2)^{D-1}D_T'(\epsilon^2)
 \over [z^{D-1} D_T(z)]'|_{z=\epsilon^2}}.
\end{eqnarray}
Hence we arrive at the expression of the nonlocal
gauge:
\begin{eqnarray}
 \eta(k^2) &=& 
 - {(D-2)\epsilon^{2(D-1)} D_T(\epsilon^2) \over
 (k^2)^{D-1} D_T(k^2)} 
{(\epsilon^2)^{D-1}D_T'(\epsilon^2)
 \over [z^{D-1} D_T(z)]'|_{z=\epsilon^2}}
 \nonumber\\&&
 - {D-2 \over (k^2)^{D-1} D_T(k^2)}
\int_{\epsilon^2}^{k^2} dz D_T'(z) z^{D-1},
\label{nlgir1}
\end{eqnarray} 
or
\begin{eqnarray}
 \eta(k^2) \equiv 1- \tilde \xi(k^2)
&=& - (D-2){\epsilon^{2(D-1)} D_T(\epsilon^2) \over
 (k^2)^{D-1} D_T(k^2)} \left[
{(\epsilon^2)^{D-1}D_T'(\epsilon^2)
 \over [z^{D-1} D_T(z)]'|_{z=\epsilon^2}} - 1 \right]
 \nonumber\\&&
 - (D-2) 
 + {(D-2)(D-1) \over (k^2)^{D-1} D_T(k^2)}
\int_{\epsilon^2}^{k^2} dz D_T(z) z^{D-2} .
\label{nlgir2}
\end{eqnarray} 
Note that 
$\tilde \xi(k^2):=1-\eta(k^2)$ is in general
different from
$\xi(k^2)$.

\section{Gauge choice and LK transformation}

It is worth remarking that the SD equation and the WT
identity are form-invariant under the  
the Landau-Khalatnikov (LK) transformation \cite{LK56}:
\begin{eqnarray}
  D_{\mu\nu}'(x) &=& D_{\mu\nu}(x) 
  + \partial_\mu  \partial_\nu 
  f(x),
  \nonumber\\
  S'(x,y) &=& e^{ e^2[f(o)-f(x-y)] } S(x,y),
  \nonumber\\
  {\cal V}_\nu'(x,y,z) 
  &=& e^{ e^2[f(o)-f(x-y)] }{\cal V}_\nu(x,y,z)
  \nonumber\\&&
  + S(x,y) e^{ e^2[f(o)-f(x-y)] } 
  \partial_\nu^z [f(x-z) - f(z-y)],
\label{LKtransf}
\end{eqnarray}
where
\begin{eqnarray}
D_{\mu\nu}(x,y) &=& \langle 0|T[A_\mu(x) A_\nu(y)]|0
\rangle,
  \nonumber\\
  S(x,y) &=& \langle 0|T[\Psi(x) \bar \Psi(y)]|0 \rangle ,
  \nonumber\\
  {\cal V}_\nu(x,y,z) 
 &=& \langle 0|T[\Psi(x) \bar \Psi(y) A_\nu(z)]|0 \rangle .
\end{eqnarray}
This can be easily shown in coordinate space where the
SD equation has the following form:
\begin{eqnarray}
 (i \hat{\partial} - m) S(x,y) 
 = \delta^D(x-y) + i e^2 \gamma^\mu 
 \langle \Psi(x) \bar \Psi(y) A_\mu(x) \rangle,
\label{cSDf}
\end{eqnarray}
and
\begin{eqnarray}
D_{\mu\nu}^{-1}(x,z)  
&=&  D_{\mu\nu}^{(0)}{}^{-1}(x,z) - \Pi_{\mu\nu}(x,z) ,
  \nonumber\\
  \Pi_{\mu\nu}(x,z) 
  &=& (g_{\mu\nu}\partial^2 - \partial_\mu \partial_\nu) 
  \Pi(x-z)
  \nonumber\\
  &=& ie^2 \int d^Dz_1 d^Dz_2 {\rm tr}[\gamma_\mu
  S(x,z_1) \Gamma_\nu(z_1,z_2;z) S(z_2,x)],
\label{cSDg}
\end{eqnarray}
where 
\begin{eqnarray}
 \langle \Psi(x) \bar \Psi(y) A_\mu(z) \rangle
 = \int d^Dx' d^Dy' d^Dz' S(x,x') \Gamma_\nu(x',y';z')
S(y',y) D_{\mu\nu}(z',z) .
\label{vertex}
\end{eqnarray}
Therefore, if we know a consistent set of solutions (for
the full gauge boson propagator, the full fermion
propagator and the full vertex function) of SD equation
in a {\it single} gauge, the solutions in other gauges
are obtained through the LK transformation.
\par
So far, the LK transformation has been used to transform
the Landau gauge result $\xi = 0$ into other gauges $\xi
\not= 0$.
Note that it is possible to perform the
inverse LK transformation (from non-Landau gauge to  
Landau gauge).  It turns out that the inverse LK
transformation is obtained from (\ref{LKtransf}) by
replacing $f(x)$ with $-f(x)$.
Furthermore, the LK transformation and its inverse
allows us to deal with the nonlocal gauge fixing, since
$f(x)$ is an arbitrary function.
Therefore, the Landau gauge result is recovered from the
nonlocal gauge by choosing
\begin{eqnarray}
 f(x) = \int {d^Dk \over (2\pi)^D} e^{i k \cdot x}
 {\xi(k) \over k^4} ,
\end{eqnarray}
where $\xi(k^2)$ is related to $\eta(k^2)$ by
(\ref{gaugerel}).
The inverse LK transformation enables us to compare the
result of this approach with the conventional one.

\par
Using the inverse LK transformation, we can obtain the
fermion propagator $S_L(x)$ in the (usual) Landau gauge
$\xi=0$ from the fermion propagator $S_{nlg}(x)$ in the
nonlocal gauge 
\begin{eqnarray}
 S_L(x) = e^{\Delta(x)} S_{nlg}(x),
 \label{LKnlg}
\end{eqnarray}
where 
\begin{eqnarray}
 \Delta(x) =  e^2 \int {d^Dk \over (2\pi)^D} 
 (e^{ik \cdot x}-1) f(k),
 \quad f(k) :=
 {\tilde \xi(k^2) \over k^4[1-\Pi(k)/k^2]} .
 \label{delta}
\end{eqnarray}
\par
In general, the fermion propagator in configuration
space is written in the form: 
\begin{eqnarray}
 \tilde S(x) = i \gamma^\mu x_\mu P(x) + Q(x),
 \label{fpc}
\end{eqnarray}
in accord with 
$S(p) = [A(p^2)\gamma^\mu p_\mu + B(p^2)]^{-1}$.
Taking into account the relation
\begin{eqnarray}
 S(p) = \int d^Dx e^{i p \cdot x} \tilde S(x)
 = {A(p^2)\gamma^\mu p_\mu + B(p^2) \over 
 A^2(p^2)p^2 + B^2(p^2)},
 \label{propa}
\end{eqnarray}
we obtain
\begin{eqnarray}
{A(p^2)p^2 \over  A^2(p^2)p^2 + B^2(p^2)}
&=& i \int d^Dx e^{i p \cdot x} (p \cdot x)P(x),
\nonumber\\
{B(p^2) \over  A^2(p^2)p^2 + B^2(p^2)}
&=& \int d^Dx e^{i p \cdot x} Q(x) .
 \label{m2}
\end{eqnarray}
Therefore, $A(p^2)$ and  $B(p^2)$ in the Landau gauge are
obtained by substituting 
$P(x)=e^{\Delta(x)} P_{nlg}(x)$
and 
$Q(x)=e^{\Delta(x)} Q_{nlg}(x)$
into (\ref{m2}).
In order to accomplish this, we must obtain the
expression of the fermion propagator 
$S_{nlg}(x)$ in configuration space.

\par
In the massless fermion phase $B(x) \equiv
0$,  the fermion propagator $S_{nlg}(p)$ in the nonlocal
gauge is nothing but the free massless propagator
$S_{nlg}(p) \equiv S_0(p) =1/(\gamma^\mu p_\mu)$.
In configuration space the free massless propagator is
given by
\begin{eqnarray}
 \tilde S_0(x) 
 = \int {d^Dp \over (2\pi)^D} e^{i p \cdot x}
 {\gamma^\mu p_\mu \over p^2}
 = i \gamma^\mu x_\mu P_0(x),
 \quad 
 P_0(x) = {\Gamma(D/2) \over 2\pi^{D/2}|x|^{D/2}}.
 \label{freeSd}
\end{eqnarray}
Therefore, the full fermion propagator in Landau gauge is
given by
\begin{eqnarray}
 S_L(p) = {\gamma^\mu p_\mu \over A(p^2)p^2},
 \quad
A^{-1}(p^2) 
= i \int d^Dx e^{i p \cdot x} e^{\Delta(x)} 
 (p \cdot x) P_0(x) .
 \label{fpL}
\end{eqnarray}
Note that the vacuum polarization  $\Pi(k)$
determines $\tilde \xi(k)$ and then $\Delta(x)$.
Therefore, the specification of the vacuum polarization is
crucial. In the quenched limit $\Pi(k) \equiv 0$,
$\tilde \xi(k) \equiv 0$ and hence 
$S_L \equiv S_0$.  This is a consistency check.
\par
In the massless fermion phase, the vacuum polarization
can be calculated easier than the massive case.
The explicit expression depends crucially on the
spacetime dimension in question.
The nonlocal gauge function 
$\tilde \xi(k)$ is expected to have a finite
range, namely, 
$|\tilde \xi(k)|<c$ uniformly in $k$, except for the
neighborhood of $k=0$.
So $\Pi(k)$ is expected not to change the UV behavior of
$f(k)$ qualitatively, see (\ref{delta}).
However, in lower dimensions $2<D<4$, $\Pi(k)$ can
influence the IR behavior of $f(k)$ considerably.
This is discussed in section 5.3 in more detail.

\section{Running coupling constant and nonlocal gauge}
\setcounter{equation}{0}

\subsection{Wavefunction renormalization and running
coupling}

Suppose that the SD equation (\ref{SDA1}) for $A$
has been solved in the Landau gauge $\eta(k^2)
\equiv 1$ under the ansatz (\ref{vertex1}) for the vertex
function  (when $B^2$ is neglected). 
Then the SD equation
(\ref{SDB}) for $B$ can be written as
\begin{eqnarray}
 B(p^2) &=& m_0 + e^2 \int {d^Dq \over (2\pi)^D} 
 {B(q^2) \over q^2} 
 \left[{G(p^2, q^2, k^2) \over A^{2}(q^2)}\right] 
 (D-1) D_T(k^2),
\label{effSD}
\end{eqnarray} 
where $B^2(q^2)$ in the denominator is neglected
according to the bifurcation method
\cite{Atkinson,Kondop}.

\par
In the paper \cite{KN92}, the following ansatz for the
vertex is adopted in order to look for the
solution in QED$_3$:  
\begin{eqnarray}
 G(p^2,q^2, k^2) = A(q^2)^n ,
 \label{ansatz3}
\end{eqnarray}
where $n$ is an integer ($n=1,2,...$).
Under this ansatz, the approximate solution
of the SD equation for $A$ was obtained in the usual
Landau gauge: 
\begin{eqnarray}
   A(p^2) = \left( 1+ {2-n \over 3} K t \right)^{1/(2-n)},
   \label{solA}
\end{eqnarray}
where
\begin{eqnarray}
  K := {8 \over \pi^2 N_f} ,
\end{eqnarray}
and
\begin{eqnarray}
  t:= \ln {p \over \alpha}, 
  \quad
  \alpha := {e^2 N_f \over 8}.
\end{eqnarray}
\par
Based on this solution, it was pointed out that one can
define a running coupling constant $K(t)$ whose actual
running is given by
\begin{eqnarray}
  K(t) :=  A_{sol}^{n-2}(p^2) K 
  = {K \over 1+{2-n \over 3} K t}  
  \label{runc}
\end{eqnarray}
due to wavefunction renormalization.
Indeed, in the absence of wavefunction renormalization
$A(p^2) \equiv 1$ and the coupling $K(t)$ does not run,
i.e. $K(t) \equiv K$. 
\par
It is claimed that the running coupling constant of
QED$_3$ obtained in such a way corresponds to the
asymptotically free case for
$n<2$ (here $n=1$ is likely to be the physical one due to
the WT identity).
If we accept Eq.~(\ref{runc}) at face value, the running
coupling
$K(t)$ diverges in the IR region for $n<2$.  In this case,
$K(t)$ becomes strong enough to be able to cause 
chiral symmetry breaking and to make the bound state
$\bar \Psi \Psi$.  It should be noted that the asymptotic
freedom defined here is valid only if 
$
t_0 := \ln {\epsilon/\alpha} 
< t_\infty:= -3/[(2-n)K].
$
Therefore this asymptotic freedom would disappear for
sufficiently large IR cutoff $\epsilon$ or small UV cutoff
$\alpha$.  This gives a possible (physical) explanation
for the controversy among the results \cite{ABKW86,PW88}
on the phase structure of QED$_3$.  For more details, see
ref.~\cite{KN92}
\par
The above observation is based on the approximate solution
(\ref{solA}) for $A$ in the Landau gauge $\xi=0$  under
the vertex ansatz (\ref{ansatz3}).
Quite recently, the rate of running of the coupling
constant $K(t)$ has been studied in more detail by
Aitchison and Mavromatos \cite{AM96} and a
subsequent paper \cite{AAKMN96} where the validity of all
the approximations made in solving the SD equation for $A$
in the paper \cite{KN92} were reexamined thoroughly under
the same vertex ansatz.

\subsection{Running coupling through nonlocal gauge}

\par
Now we study the running coupling based on the nonlocal
gauge. In the nonlocal gauge, the function $B$ obeys the
SD equation (\ref{SDBnlg}).
If we choose the nonlocal gauge $\xi(k)$ such that 
$A(p^2) \equiv 1$, the function $G(p^2, q^2, k^2)$ should
be replaced with 1, i.e. $G(p^2, q^2, k^2) \equiv 1$ from
the consistency with the WT identity. In this case, the
wavefunction renormalization disappears and the SD
equation for $B$ reduces to
\begin{eqnarray}
 B(p^2) &=& m_0 + e^2 \int {d^Dq \over (2\pi)^D}
  {B(q^2) \over q^2 +B^2(q^2)} [D-\eta(k^2)] D_T(k^2) .
\label{SDBnlg3}
\end{eqnarray} 
This should be compared with the SD equation (\ref{effSD}).
Then the running coupling constant is given by
\begin{eqnarray}
   K(t)/K = {D-\eta(k^2)\over D-1}  
   = {D-1+\tilde \xi(k^2)\over D-1}  .
   \label{runcoup}
\end{eqnarray} 
Here note that the argument of the running coupling
constant is the gauge-boson momentum, which is
required for the nonlocal gauge to be consistent with
the axial WT identity as well as the vector WT identity
\cite{KM92}.
Therefore, the problem of finding the running coupling
constant reduces to finding the nonlocal gauge.  

Therefore we can see the following advantage or benefits
of this approach:

\begin{enumerate}
\item[1)]
We do not have to assume a particular ansatz for the
vertex.  We can study a class of vertices of the
form:
$
 \Gamma_\mu(p,q) = \gamma_\mu G(p^2,q^2, k^2),
$
without assuming any particular form of $G(p^2,q^2,
k^2)$. The consistency with the WT identity requires that 
$G(p^2,q^2, k^2) \rightarrow 1$ as $A(p^2) \rightarrow
1$.  This class of vertex ansatz includes the previous one
(\ref{ansatz3}) with $n=1$.

\item[2)]
Under this ansatz for the vertex, we need not
to solve the SD equation for $A(p^2)$. 
This releases us from worrying about the validity of a
number of approximations which are required to solve the
SD equation for $A$.
Instead, we obtain the differential equation for the
nonlocal gauge.  This is solved by simple quadrature.
In particular, we do not have to perform any angular
integration.  So  the Higashijima-like approximation
\cite{HM84} is unnecessary to separate the kernel
of the integral equation (at least in the normal phase).

\item[3)]
If necessary, we can recover the
solution in the usual gauge  (e.g. Landau gauge), by
making use of the (inverse) LK transformation as
discussed in the previous section. This enables us to
compare the result in this approach with the conventional
result.

\item[4)]
We can study the effect of a cutoff in more detail.

\end{enumerate}

The final point needs more explanations.
In this approach it is not necessary to introduce the UV
cutoff $\alpha$ in QED$_3$.
In the analysis of QED$_3$ \cite{KN92}, the existence
of the IR cutoff $\epsilon$ was essential as well
as the UV cutoff $\alpha$.
Whether or not there exists a "finite" critical number
of flavors $N_f^c$ above which ($N_f >N_f^c$) 
spontaneous chiral symmetry breaking disappear depends
crucially on the existence of the infrared cutoff and the
ratio $\epsilon/\alpha$.
It is expected that the asymptotic freedom claimed above
would disappear for sufficiently large IR cutoff
$\epsilon$ or small UV cutoff
$\alpha$ and hence this leads to a finite critical number
of flavors $N_f^c<\infty$ for sufficiently large
$\epsilon/\alpha$. This observation is based on a naive
analogy with QCD in 3+1 dimensions (QCD$_4$). In QCD$_4$, the
theory has only one phase, the chiral-symmetry breaking
and confining phase, due to asymptotic freedom where the
running of the gauge coupling constant is essential
\cite{HM84}. On the other hand, QED$_4$ is not
asymptotically free and has a critical gauge coupling
constant $e_c$ above which
($e>e_c$) a fermion mass is dynamically generated and
 chiral symmetry is spontaneously broken. Therefore,
the IR behavior of the running coupling constant in QED3
is very important to resolve the phase structure of
QED$_3$.
\par
On the the hand, QED$_3$ has  already been
studied in the nonlocal gauge \cite{KEIT95,KM95}
where we have found a phase transition at a finite
critical number of flavors
$N_f^c<\infty$ which separates
the chiral-symmetric phase from the
spontaneous-chiral-symmetry-breaking phase. In this
analysis we did not introduce the infrared cutoff
from the beginning. Apparently, the results
of \cite{KN92} and
\cite{KEIT95} contradict with each other, since the very
small IR cutoff (compared with UV cutoff $\alpha$) should
lead to the asymptotic freedom and no phase
transition, i.e. non-existence of a finite critical
number of flavors according to \cite{KN92}. 
In this paper we resolve this apparent contradiction.
For this, we introduce the infrared cutoff in the setting
up of nonlocal gauge. 
The result is given in the next sections.

\subsection{Inverse LK transformation and wavefunction
renormalization}

\par
Thanks to the inverse LK transformation, we can obtain
the fermion propagator $S_L(x)$ in the (usual) Landau
gauge $\xi=0$ from the fermion propagator $S_{nlg}(x)$ in
the nonlocal gauge.
In $D=3$ dimensions, the free massless propagator in the
configuration space is given by
\begin{eqnarray}
 S_0(x) = {\gamma^\mu x_\mu \over 4\pi|x|^3}.
 \label{freeS3d}
\end{eqnarray}
For $D=3$, the massless fermion generates the
vacuum polarization (at one-loop):
\begin{eqnarray}
\Pi(k) = - \alpha k .
\label{vp3}
\end{eqnarray}
\par
By substituting the vacuum polarization function
(\ref{vp3}) and
$S_{nlg}(x)=S_0(x)$ given by (\ref{freeS3d}) 
into (\ref{LKnlg}) with the nonlocal gauge function 
$\tilde \xi(k^2)$ obtained explicitly in the next
section, we can get the fermion propagator $S_L(x)$ in the
Landau gauge in configuration space.
\par
If $\xi$ was a constant, we would have for $D=3$
\begin{eqnarray}
 \Delta(x) =  e^2 \int {d^3k \over (2\pi)^3} 
 (e^{ik \cdot x}-1)  {\xi \over k^4}
 = - {e^2 \over 8\pi} \xi |x|.
\end{eqnarray}
According to (\ref{fpL}), it is not difficult to show
that the wavefunction renormalization function is
obtained  as 
\begin{eqnarray}
 A^{-1}(p^2) =  1-{e^2 \xi \over 8\pi p} 
 \arctan ({8\pi p \over e^2 \xi})  .
\end{eqnarray}
In the IR region, we find that
\begin{eqnarray}
 A^{-1}(p^2) 
 =  {1 \over 3}({8\pi p \over e^2 \xi})^2  + O(p^4) .
\end{eqnarray}
This is totally different from what we expect based on the
nonlocal gauge.
\par
In the nonlocal gauge,  the
integrand $f(k)$ of
$\Delta(x)$ in the UV region $k \rightarrow
\infty$ exhibits the behavior
\begin{eqnarray}
 f(k) \sim  {\tilde \xi(k^2) \over k^4} \rightarrow 0 
 \quad (k \rightarrow \infty),
\end{eqnarray}
so the effect of $\Pi$ is neglected in this region. 
However, in the IR region 
$k \rightarrow 0$,
the presence of $\Pi(k)$ totally changes the situation:
\begin{eqnarray}
 f(k) \sim  {\tilde \xi(k^2) \over \alpha k^3}
 \quad (k \rightarrow 0).
\end{eqnarray}
In either case, the situation does not resemble the
constant $\xi$ case.
If the integration is performed numerically in
(\ref{fpL}), we will be able to obtain the wavefunction
renormalization function
$A(p^2)$ in the Landau gauge (the corresponding vertex
function can be also obtained from the LK transformation).
The numerical result will be given elsewhere.

\section{Running coupling constant of QED$_3$}
\setcounter{equation}{0}
\par
In $D=3$ dimensions, the running coupling (\ref{runcoup})
is obtained by shifting and scaling the nonlocal gauge
according to
\begin{eqnarray}
   K(t)/K = 1 + {\tilde \xi(k^2) \over 2}  .
   \label{runc3d}
\end{eqnarray} 
The exact non-trivial wavefunction renormalization in
the Landau gauge is obtained from (\ref{fpL}).
Under a specific ansatz (\ref{ansatz3}), it is obtained
from  (\ref{runc}) and (\ref{runcoup}) as
\begin{eqnarray}
   A(p^2) 
 = \left( 1 + {\tilde \xi(p^2) \over 2}\right)^{1/(n-2)},
\end{eqnarray} 
and in particular for $n=1$
\begin{eqnarray}
   A(p^2) 
 = \left( 1 + {\tilde \xi(p^2) \over 2}\right)^{-1}.
 \label{A3d}
\end{eqnarray} 
Therefore, we study the behavior of the nonlocal gauge
function $\tilde \xi(k^2)$ in what follows.

\subsection{Running coupling constant (I)}

From Eq.~(\ref{nlgir2}), the nonlocal gauge of QED$_3$
with an IR cutoff $\epsilon$ is given by
\begin{eqnarray}
 \tilde \xi_\epsilon(k^2) &=& 
 \xi_\epsilon^a(k^2) + \xi_\epsilon^b(k^2),
 \nonumber\\
\xi_\epsilon^a(k^2) &:=&  2
 - {2 \over (k^2)^{2} D_T(k^2)}
\int_{\epsilon^2}^{k^2} dz D_T(z) z ,
 \nonumber\\
\xi_\epsilon^b(k^2) &:=&  {\epsilon^{4} D_T(\epsilon^2)
\over  (k^2)^{2} D_T(k^2)} \left[
{(\epsilon^2)^{2}D_T'(\epsilon^2)
 \over [z^{2} D_T(z)]'|_{z=\epsilon^2}} - 1 \right] .
\label{nlgirqed3I}
\end{eqnarray} 
Here we decomposed the nonlocal gauge into two pieces:
the first piece $\xi_\epsilon^a$ reduces in the limit 
$\epsilon \rightarrow 0$ to the nonlocal
gauge without an IR cutoff, while the second piece
$\xi^b$ comes from the flatness condition
$\xi'(\epsilon^2)=0$ and vanishes in the limit 
$\epsilon \rightarrow 0$.
 First of all, we consider the following integral
\begin{eqnarray}
  J_1(k^2; \epsilon) &:=& \int_{\epsilon^2}^{k^2} dz 
  {z \over z + \alpha \sqrt{z}}
  \nonumber\\
  &=& k^2 - 2 \alpha \sqrt{k^2}
  - \epsilon^2 + 2 \alpha \sqrt{\epsilon^2} 
  + 2 \alpha^2 \ln {k+\alpha 
  \over \sqrt{\epsilon^2}+\alpha} .
\end{eqnarray}
This has the expansion:
\begin{eqnarray}
J_1(k^2;\epsilon)
=(-{{\epsilon}^2} + 2\,\alpha \,{\sqrt{{{\epsilon}^2}}} + 
    2\,{{\alpha }^2}\,\log (\alpha ) - 
    2\,{{\alpha }^2}\,\log (\alpha  +
{\sqrt{{{\epsilon}^2}}})) + 
  {{2\,{k^3}}\over {3\,\alpha }} + O(k^4) .
\end{eqnarray}
Hence the first piece $\xi^a$ of the nonlocal gauge
(\ref{nlgirqed3I}) given by
\begin{eqnarray}
  \xi_\epsilon^a(k^2) 
  = 2 - 2 {k^2+\alpha \sqrt{k^2} \over (k^2)^2}
J_1(k^2;\epsilon^2)  
\end{eqnarray}
has the following expansion around $k=0$:
\begin{eqnarray}
  \xi_\epsilon^a(k^2) 
&=& 
{{2\,\alpha \,\left( {{\epsilon }^2} - 
        2\,\alpha \,{\sqrt{{{\epsilon }^2}}}  + 
        2\,{{\alpha }^2}\,
         \log (1  + {\sqrt{{{\epsilon }^2}}}/\alpha) \right)
}\over  {{k^3}}} 
\nonumber\\&&
      + {{2 \,\left( {{\epsilon }^2} - 
        2\,\alpha \,{\sqrt{{{\epsilon }^2}}}  + 
        2\,{{\alpha }^2}\,
         \log (1  + {\sqrt{{{\epsilon }^2}}}/\alpha) \right)
}\over  {{k^2}}} 
\nonumber\\&&
      + {2\over 3} - 
  {k\over {3\,\alpha }} + 
  {{{k^2}}\over {5\,{{\alpha }^2}}} - 
  {{2\,{k^3}}\over {15\,{{\alpha }^3}}} + O(k^4) .
  \label{xia}
\end{eqnarray}
Note that the introduction of the IR cutoff $\epsilon$ 
generates  IR singular terms like $2\alpha c k^{-3}$, 
$2ck^{-2}$ where the coefficient $c$ is always positive. 
Such a singular behavior in the IR region disappears if we
put $\epsilon=0$ from the beginning, and $\xi^a$ reduces
to
\begin{eqnarray}
  \tilde \xi_0(k^2) 
= {2\over 3} - 
  {k\over {3\,\alpha }} + 
  {{{k^2}}\over {5\,{{\alpha }^2}}} - 
  {{2\,{k^3}}\over {15\,{{\alpha }^3}}} + O(k^4) .
  \label{nlgQED3}
\end{eqnarray}
This is nothing but the result obtained in \cite{KEIT95} 
(by setting the Chern-Simons coefficient
$\theta$ equal to zero: $\theta=0$ in eq.~(29) of
\cite{KEIT95}). On the other hand, the second piece is
always negative and singular at $k=0$:
\begin{eqnarray}
\xi_\epsilon^b(k^2) =
- {{4\,\left( {k^2} + \alpha \,{\sqrt{{k^2}}}
\right)
\,{{\epsilon }^3}\, }\over 
{{k^4}\,\left( 3\,\alpha  + 2\,{\sqrt{{{\epsilon }^2}}}
\right)
\, }}
= - { 4 \epsilon ^3 \over 
      \left( 3 \alpha  + 2 \sqrt{\epsilon^2} \right)}
      \left( {\alpha \over k^3} + {1 \over k^2} \right) .
      \label{xib}
\end{eqnarray}
By adding (\ref{xia}) and (\ref{xib}), we find that the
singular part with negative power of $k$ in the nonlocal
gauge is negative.
\par
In the region  $k/\alpha \gg 1$, $\xi^a$ is dominant,
because $\xi^b$ decreases more rapidly than
$\xi^a$ which behaves as
\begin{eqnarray}
  \xi_\epsilon^a(k^2) 
&=& 
{{2\,\alpha }\over k} + {{4\,{{\alpha }^2} + 2\,{{\epsilon }^2} - 
      4\,\alpha \,{\sqrt{{{\epsilon }^2}}} 
    + 4\,{{\alpha }^2}\,
      \log ((\alpha  + {\sqrt{{{\epsilon }^2}}})/k)}
     \over {{k^2}}} 
    + O({1\over k})^3 .
\end{eqnarray}
It turns out that the $\xi_\epsilon^a$ alone
is monotonically decreasing in
$k$ with a maximum value $2$ at $k=\epsilon$ (and
diverges monotonically as $k \rightarrow 0$) which rather
enhances the effective coupling 
compared with $\xi_0(0)=2/3$.
This tendency agrees with the observation made in the
previous paper \cite{KN92} where the running of
the coupling is terminated at $k=\epsilon$ and the
flatness of the coupling  
$K(k^2)=K(\epsilon^2)$ for $k \le \epsilon$ is assumed a
priori, which is borrowed from the QCD$_4$ analysis. 
If we consider only the $\xi^a$ piece,
there is a discontinuity in the derivative
$\xi_\epsilon^a{}'(k)$ at $k=\epsilon$, since
$\xi_\epsilon^a{}'(\epsilon+0)<0$ and
$\xi_\epsilon^a{}'(\epsilon-0)=0$.
However, the inclusion of 
$\xi_\epsilon^b(<0)$ which is necessary to satisfy
$\tilde \xi_\epsilon'(\epsilon)=0$ (continuity of
$\tilde \xi_\epsilon'(k)$ at $k=\epsilon$) considerably
changes the situation.
The inclusion of such a term causes a slowing down of
the rate of decrease of the effective coupling constant
$K(t)$.  This tendency agrees with the recent analysis of
\cite{AAKMN96}. The  the nonlocal gauge function and the
running coupling constant are monotonically decreasing in
$k$ for
$k>\epsilon$ and have upper bonds:
\begin{eqnarray}
\tilde \xi_\epsilon(\epsilon^2) 
=\xi_\epsilon^a(\epsilon^2) + \xi_\epsilon^b(\epsilon^2) 
= 2-{4(\alpha+\epsilon) \over 3\alpha+2\epsilon}
= {2 \over 3+2\epsilon/\alpha}
< {2 \over 3}.
\end{eqnarray}
For $k<\epsilon$, the nonlocal gauge $\tilde
\xi_\epsilon$ decreases as
$k$ decreases.
In Figure 1, the nonlocal gauge is plotted for  
$\epsilon/\alpha = 0.1$.
The running coupling constant is obtained from
(\ref{runc3d}), and the wavefunction renormalization in
the Landau gauge is obtained from (\ref{A3d}).

\subsection{Running coupling constant (II)}

In this section, we consider another way of introducing
the IR cutoff.  We introduce the IR cutoff $\delta$ in
the gauge-boson propagator:
\begin{eqnarray}
  D_T(k^2) = {1 \over k^2 + \alpha k + \delta^2}.
\end{eqnarray}
The cutoff $\delta$ plays the same role as the
gauge-boson mass and seems to be more natural than the
previous one.  
This choice of $D_T$ is equivalent to the gauged Thirring
model in the nonlocal $R_\xi$ gauge \cite{Kondo96a}
where $\delta^2=e^2 G_T^{-1}$ for the Thirring coupling
$G_T$.
\par
In this case, the nonlocal gauge reads
\begin{eqnarray}
  \tilde \xi_\delta(k^2) 
  = 2 - 2 {k^2+ \alpha k +\delta^2 \over (k^2)^2}
J_2(k^2,\delta), 
\label{nlgirqed3II}
\end{eqnarray}
with
\begin{eqnarray}
  J_2(k^2; \delta) &:=& \int_{0}^{k^2} dz 
  {z \over z + \alpha \sqrt{z}+ \delta^2}
  \nonumber\\
  &=& \int_{0}^{k} dr 
  {2r^3 \over r^2 + \alpha r+ \delta^2}
  \nonumber\\
  &=&    (k -2 \alpha)k
  + (\alpha^2 - \delta^2)
  \ln |(k^2+\alpha k +\delta^2)/\delta^2|
  \nonumber\\&&
  + \alpha (3\delta^2 - \alpha^2) [I_B(k)- I_B(0)] ,
\end{eqnarray}
where the indefinite integral $I_B(k)$ is defined by
\begin{eqnarray}
  I_B(k)  := \int^{k} dr 
  {1 \over r^2+\alpha r+\delta^2} .
\end{eqnarray}
The indefinite integral  $I_B(k)$ can be calculated: For
$\alpha^2>4\delta^2$, 
\begin{eqnarray}
  I_B(k)  =  {-1 \over \sqrt{\alpha^2-4\delta^2}}
  \ln {(2k+\alpha+\sqrt{\alpha^2-4\delta^2})^2
\over |k^2+\alpha k+\delta^2|} ,
\end{eqnarray}
and for $\alpha^2<4\delta^2$, 
\begin{eqnarray}
  I_B(k)  =  {2 \over \sqrt{4\delta^2-\alpha^2}}
  \arctan {2k+\alpha \over \sqrt{4\delta^2-\alpha^2}} .
\end{eqnarray}
\par
When $\alpha^2>4\delta^2$, a careful analysis shows
that all the terms in $J_2$ up to $O(k^3)$ cancel,
namely,
\begin{eqnarray}
  J_2(k^2; \delta) 
  = {1 \over 2\delta^2} k^4 
  - {2\alpha \over 5\delta^4}k^5
  + {\alpha^2-\delta^2 \over 3\delta^6} k^6 
  + O(k^7) .
\end{eqnarray}
This implies the following expansion of the nonlocal
gauge $\tilde \xi_\delta(k^2)$ around $k=0$:
\begin{eqnarray}
  \tilde \xi_\delta(k^2) 
  = 1 - {\alpha \over 5\delta^2} k  
  + {2\alpha^2-5\delta^2 \over 15\delta^4} k^2 
  - {2\alpha(5\alpha^2-17\delta^2) \over 105 \delta^6} k^3
  + O(k^4) .
  \label{nonlocalgth}
\end{eqnarray}
This result shows that there is no singularity in the
nonlocal gauge (\ref{nlgirqed3II}) at $k=0$, in sharp
contrast with the nonlocal gauge $\tilde
\xi_\epsilon(k^2)$. In the region
$k/\alpha \ll 1$,  the nonlocal gauge behaves as
follows:
at $\delta/\alpha=0.1$
\begin{eqnarray}
  \tilde \xi_\delta(k^2) 
=   1 -  20 (k/\alpha) + 1300  (k/\alpha)^2 
- 92000 (k/\alpha)^3
+  O((k/\alpha)^4) ,
\end{eqnarray}
and at $\delta/\alpha=0.01$
\begin{eqnarray}
  \tilde \xi_\delta(k^2) 
&=&  1 -  2000 (k/\alpha) 
+ 1.333 \times {10}^7 (k/\alpha)^2  
- 9.52057 \times 10^{10} (k/\alpha)^3
\nonumber\\&&
+   O((k/\alpha)^4) .
\end{eqnarray}

\par
It should be remarked that, in the limit $(\delta
\rightarrow 0)$, the nonlocal gauge (\ref{nonlocalgth})
does not reduce to the expected form  (\ref{nlgQED3}).
There is a discontinuity between the nonlocal gauge
$\tilde \xi_\delta(k^2)$ with an IR cutoff $\delta$ and
the nonlocal gauge $\tilde \xi_0(k^2)$ obtained
without an IR cutoff from the beginning. 
Such a discontinuity does not exist for the nonlocal
gauge $\tilde \xi_\epsilon(k^2)$ with IR cutoff
$\epsilon$.
Such a situation can be seen in the finite temperature
case: there may occur some discontinuity between the
zero-temperature limit of finite-temperature calculation
and the corresponding result evaluated at
zero-temperature. In this context $\delta$ can be
interpreted as a plasmon mass. This discontinuity is very
similar to the result found by Aitchison et al.
\cite{AAKMN96}. The non-local gauge $\tilde
\xi_\delta(k^2)$ decreases more slowly than claimed in
the previous paper
\cite{KN92}. 
The nonlocal gauge is plotted in Figure 2.

\subsection{RG-like point of view}

By using the idea of RG, we show
that the IR and UV behaviors of the nonlocal gauge
$\tilde \xi_\delta(k^2)$ and the effective running
coupling constant can be easily analyzed without
performing any integration.  For this, we calculate the
derivative of the nonlocal gauge function and express the
result in terms of the nonlocal gauge function itself:
\begin{eqnarray}
  \beta(\tilde \xi) 
:= -  {d \over d \ln (k/\mu)} \tilde \xi(k^2) 
=    4 + {2k^2+3\alpha k+4\delta^2 \over 
k^2+\alpha k+\delta^2} (\tilde \xi(k^2)-2).
\end{eqnarray}
From this equation, we can observe the following:
\par\noindent
1) without IR cutoff ($\delta=0$), in the IR limit $k
\rightarrow 0$ 
\begin{eqnarray}
  \beta(\tilde \xi)
=    4 + 3(\tilde \xi(k^2)-2) =  3 \tilde \xi(k^2) - 2.
\end{eqnarray}
This implies 
$
 \tilde \xi(k^2) \rightarrow 2/3 = \tilde \xi_0(0)
$
in the IR limit $k\rightarrow 0$.

\par\noindent
2) with IR cutoff ($\delta\not=0$), in the IR limit $k
\rightarrow 0$ 
\begin{eqnarray}
  \beta(\tilde \xi)
=  4 + 4(\tilde \xi(k^2)-2) =  4 \tilde \xi(k^2) - 4.
\end{eqnarray}
This implies 
$
\tilde \xi(k^2) \rightarrow 1
$ 
in the IR limit $k\rightarrow 0$.
\par
These result seems to show the existence of non-trivial
IR fixed point for the running coupling constant
$\bar K:=K(t)$ defined by 
$
K(t)/K = 1+\tilde \xi(k^2)/2 
$
at $\bar K=4/3 K$ and $\bar K=3/2 K \sim O(1/N)$
corresponding to 1) and 2) respectively.
\par
In particular,
\par\noindent
3) in the UV limit $k \rightarrow \infty$
(irrespective of the value of the IR cutoff), 
\begin{eqnarray}
  \beta(\tilde \xi)
=  4 + 2(\tilde \xi(k^2)-2) =  2 \tilde \xi(k^2) .
\end{eqnarray}
This implies 
$
\tilde \xi(k^2) \rightarrow 0
$ 
in the UV limit $k\rightarrow \infty$.

Especially,
\par\noindent
4) in the quenched limit ($N_f \rightarrow 0$ and $\delta
= 0$),
\begin{eqnarray}
  \beta(\tilde \xi)
=   4 + 2(\tilde \xi(k^2)-2) =  2 \tilde \xi(k^2) .
\end{eqnarray}
for any $k$.
This shows that the constant solution is possible:
$\tilde \xi(k^2) \equiv 0$.

\par
It is straightforward to generalize the above argument to
arbitrary dimension.

\section{Conclusion and discussion}
\setcounter{equation}{0}

In this paper we have discussed an alternative approach
for obtaining the effective or running coupling constant
and the RG property of gauge theory in the SD framework,
which is appropriate for non-perturbative study of gauge
theory. This approach can be applied only to a gauge
theory, as we use the the gauge invariance as an essential
ingredient in the approach.
We argued that this approach is superior in
several respects to the previous conventional approach
in which the SD integral equation has been solved to
obtain the wavefunction renormalization function
$A(p^2)$ under a specific vertex ansatz to define the
running coupling. In this approach we do not solve the SD
equation for
$A$, instead we obtain the differential equation which the
nonlocal gauge function must satisfy in order that $A(p^2)
\equiv 1$.
\par
The validity of this approach has been exemplified in the
study of the IR behavior of the running coupling constant
and non-trivial IR fixed point in QED$_3$. This approach
confirms a recent result:
the slowing down of the rate of decrease of the running
coupling constant and the existence of non-trivial IR
fixed point, as claimed in \cite{AM96,AAKMN96},
but here without relying on  a specific ansatz of vertex
and on a number of approximations adopted to solve the SD
equation in the conventional approach \cite{KN92}.
\par
In this paper we have not studied the dynamical mass
generation and spontaneous chiral symmetry breaking
\cite{ABKW86} in QED$_3$ by solving the SD equation for
$B$ in the nonlocal gauge. The nonlocal gauge $\tilde
\xi_\delta(k^2)$ does not qualitatively change the
previous result \cite{KEIT95} in the nonlocal gauge
$\tilde \xi_0(k^2)$ without IR cutoff. 
For example, there should exist a certain finite
critical number of flavors $N_f^c$ above which  
dynamical mass is not generated and chiral symmetry
is restored. 
\par
The relationship between our approach and
the conventional one can be seen by using the LK
transformation.  Under  an LK transformation, the SD
equation and the WT identity are form-invariant.
By making use of an LK transformation, the non-trivial
wavefunction renormalization
$A(p^2)$ in the Landau gauge can be obtained from the
nonlocal gauge, as shown in section 4.3.
Thus the conventional picture can be recovered by the
(inverse) LK transformation from our approach, if
desired.
It would be interesting to 
apply this procedure to the case of QED$_3$ for which
Maris \cite{Maris96} has recently given a rather full
discussion of the SD equation in the conventional
framework.

\par
The nonlocal gauge has been applied to gauge
theories in four dimensions \cite{GSC90,KM92}, QED$_3$
\cite{Simmons90,KEIT95,KM95} and the gauged Thirring model
\cite{Kondo96a} under a bare vertex approximation.
Obviously, this approach is not so systematic as the
perturbative method.  The nonlocal gauge must be obtained
case by case and there is no guarantee that such a gauge
does exist beyond this order of truncation of SD
equation. 
Nevertheless it is sufficiently interesting to warrant
the extension of this approach to finite
temperature gauge field theory \cite{FTQED}.
In such a case, the nonlocal gauge may only be obtained
approximately, as in the case for QED$_3$ with a
Chern-Simons term
\cite{KEIT95,KM95}.

\section*{Acknowledgments}
The author would like to thank Prof. Ian J.R. Aitchison for
kind hospitality in Oxford and helpful discussions.
He is also grateful to Drs. N. Mavromatos and G.
Amelino-Camelia for discussions.
This work is supported in part by the Japan
Society for the Promotion of Science and the Grant-in-Aid
for Scientific Research from the Ministry of Education,
Science and Culture (No.07640377).

\centerline{{\large Figure Captions}}

\begin{enumerate}
\item[Fig.1:]
Plot of nonlocal gauge given by (\ref{nlgirqed3I}) as a
function of $k/\alpha$. 
Three graphs correspond to $\xi^a(k)$ (above),
$\tilde \xi_0(k)$ (middle) and
$\tilde \xi_\epsilon(k)$ (below).
Here we have chosen $\epsilon/\alpha=0.1$

\item[Fig.2:]
Plot of nonlocal gauge given by (\ref{nlgirqed3II}) as a
function of $k/\alpha$. 
Two graph correspond to 
$\epsilon/\alpha=0.1$ (above)
and $\epsilon/\alpha=0.01$ (below).

\end{enumerate}

\end{document}